\documentclass[%
showpacs,
nofootinbib,
 twocolumn, amsmath,amssymb,prl,aps
]{revtex4-1}

\usepackage{graphicx}
\usepackage{dcolumn}
\usepackage{bm}
\usepackage{subfigure} 

\def\bea{\begin{eqnarray}}
\def\eea{\end{eqnarray}}
\def\J{\text{J}}

\newcommand{\sgn}{\mathrm{sgn}}

\newcommand{\rd}{\mathrm{d}}
\newcommand{\bo}{\hat{b}^{\phantom\dag}}
\newcommand{\ba}{\hat{b}^{\dag}}
\renewcommand{\ao}{\hat{a}^{\phantom\dag}}
\renewcommand{\aa}{\hat{a}^{\dag}}
\newcommand{\no}{\hat{n}}
\newcommand{\Ho}{\hat{H}}
\newcommand{\Uo}{\hat{U}}
\newcommand{\la}{\langle}
\newcommand{\ra}{\rangle}
\newcommand{\be}{\begin{equation}}
\newcommand{\ee}{\end{equation}}
\newcommand{\bes}{\begin{eqnarray}}
\newcommand{\ees}{\end{eqnarray}}

\begin{document}

\title{Floquet realization and signatures of one-dimensional anyons in an 
optical lattice}

\author{Christoph Str\"ater${}^1$, Shashi C.\ L.\ Srivastava${}^{1,2}$, Andr\'e Eckardt${}^1$}

\email{eckardt@pks.mpg.de}

\affiliation{$^1$Max-Planck-Institut f\"ur Physik komplexer Systeme, N\"othnitzer Str.\ 38, 01187 Dresden, Germany}

\affiliation{$^2$Variable Energy Cyclotron Centre, 1/AF Bidhan nagar, Kolkata, India 700 064}

\begin{abstract}
We propose a simple scheme for mimicking the physics of one-dimensional anyons in an 
optical-lattice experiment. It relies on a bosonic representation of the anyonic Hubbard 
model to be realized via lattice-shaking-induced resonant tunneling against potential
off-sets created by a lattice tilt and strong on-site interactions. No lasers additional 
to those used for the creation of the optical lattice are required. We also discuss
experimental signatures of the continuous interpolation between bosons and fermions when
the statistical angle $\theta$ is varied from 0 to $\pi$. Whereas the real-space density 
of the bosonic atoms corresponds directly to that of the simulated anyonic model, this is 
not the case for the momentum distribution. Therefore, we propose to use Friedel 
oscillations in the density as a probe for continuous fermionization of the bosonic atoms.
\end{abstract}

\maketitle


Fundamental particles in nature are either bosons or fermions. Bosons obey Bose-Einstein 
statistics such that their joint wavefunction is symmetric with resepct to the exchange 
of two particles, whereas fermions obey Fermi-Dirac statistics and the wave function 
picks up a minus sign under particle exchange. In two dimensions, also anyons would be 
possible fundamental particles. They obey a fractional statistics that interpolates 
between bosonic and fermionic behavior \cite{LeinaasEtAl77, GoldinEtAl81, Wilczek82,
TsuiEtAl82, CanrightEtAl89}. If two anyons exchange their position, the wave function pics 
up a phase. Practically, anyons play a major role as quasiparticles of topologically ordered 
states of matter such as fractional-quantum-Hall states \cite{Laughlin83, Halperin84, 
CaminoEtAl05}, with potential applications in robust topological quantum information 
processing \cite{Kitaev03, DasSarmaEtAl05, BondersonEtAl06, SternEtAl06, NayakEtAl08,
AliceaEtAl11,SternEtAl13,MatthewsEtAl13}. 
As shown by Haldane, for quasiparticles the concept of fractional statistics can be 
extended to arbitrary dimensions \cite{Haldane91}. One-dimensional (1D) anyons have 
recently attracted an increased attention \cite{Ha94,MurthyEtAl94,WuEtAl95,ZhuEtAl96,
AmicoEtAl98, KunduEtAl99,BatchelorEtAl06,Girardeau06,CalabreseEtAl07,delCampo08,HaoEtAl08,
HaoEtAl09, HaoEtAl12, WanWanZha2014, TangEtAl15, ZhangEtAl15}, including two proposals 
for their implementation with bosonic atoms in an optical lattice \cite{KeilmannEtAl11,
GreschnerSantos15}. These proposals are based on mapping the anyons via a generalized
Jordan-Wigner transformation to bosons with a density-dependent tunneling parameter 
to be engineered by laser-dressing of internal atomic degrees of freedom. However, 
an experimental realization has not yet been achieved. 

In the following, we propose a simple alternative scheme for the experimental realization 
of 1D anyons, based on time-periodic forcing. It is feasible in existing experimental 
setups and, in contrast to earlier proposals, does neither rely on the internal 
atomic structure nor require any lasers additional to those creating the optical lattice. 
Our scheme is based on engineering an occupation-dependent Peierls phase of the tunneling 
matrix elements by means of coherent lattice-shaking-assisted tunneling against potential offsets, which are created by a combination of a lattice tilt and strong on-site interactions.. The scheme, which is applicable in the low-density regime, also permits to 
effectively tune the interactions between the anyons. The fact that periodic forcing has 
recently been employed already experimentally for engineering both number-dependendent 
tunneling amplitudes \cite{MaEtAl11,MeinertEtAl16} and non-number-dependent Peierls 
phases \cite{AidelsburgerEtAl11, StruckEtAl11, StruckEtAl12, StruckEtal13, 
AidelsburgerEtAl13,  MiyakeEtAl13,  AtalaEtAl14, JotzuEtAl14, 
AidelsburgerEtAl15, KennedyEtAl15} (see also Ref.~\cite{Eckardt16} for an overview of 
Floquet engineering in optical lattices) indicates that the proposed creation of
number-dependent Peierls phases by such means is feasible. 

We, moreover, discuss experimental signatures 
of the anyonic model in its ground state using exact 
diagonalization. Considering small chains, as they can be realized in quantum-gas 
microscopes \cite{IslamEtAl15,FukuharaEtAl15}, we monitor experimentally measurable 
observables that directly reflect anyonic properties and are invariant under the Jordan-
Wigner transformation from anyons to bosons. This excludes the momentum distribution, 
which is altered by the transformation so that the measurable bosonic momentum 
distribution does not correspond to that of the anyons. It includes, however, on-site 
densities and their correlations, as well as the second R\'enyi entropy of the subsystem 
given by the first $\ell$ sites. We show that in small systems Friedel oscillations can 
serve as a signature for the continuous fermionization occurring when the statistical 
angle $\theta$ is varied from $0$ to $\pi$.

\begin{figure}[t]
\centering
\includegraphics[width=0.5\textwidth,]{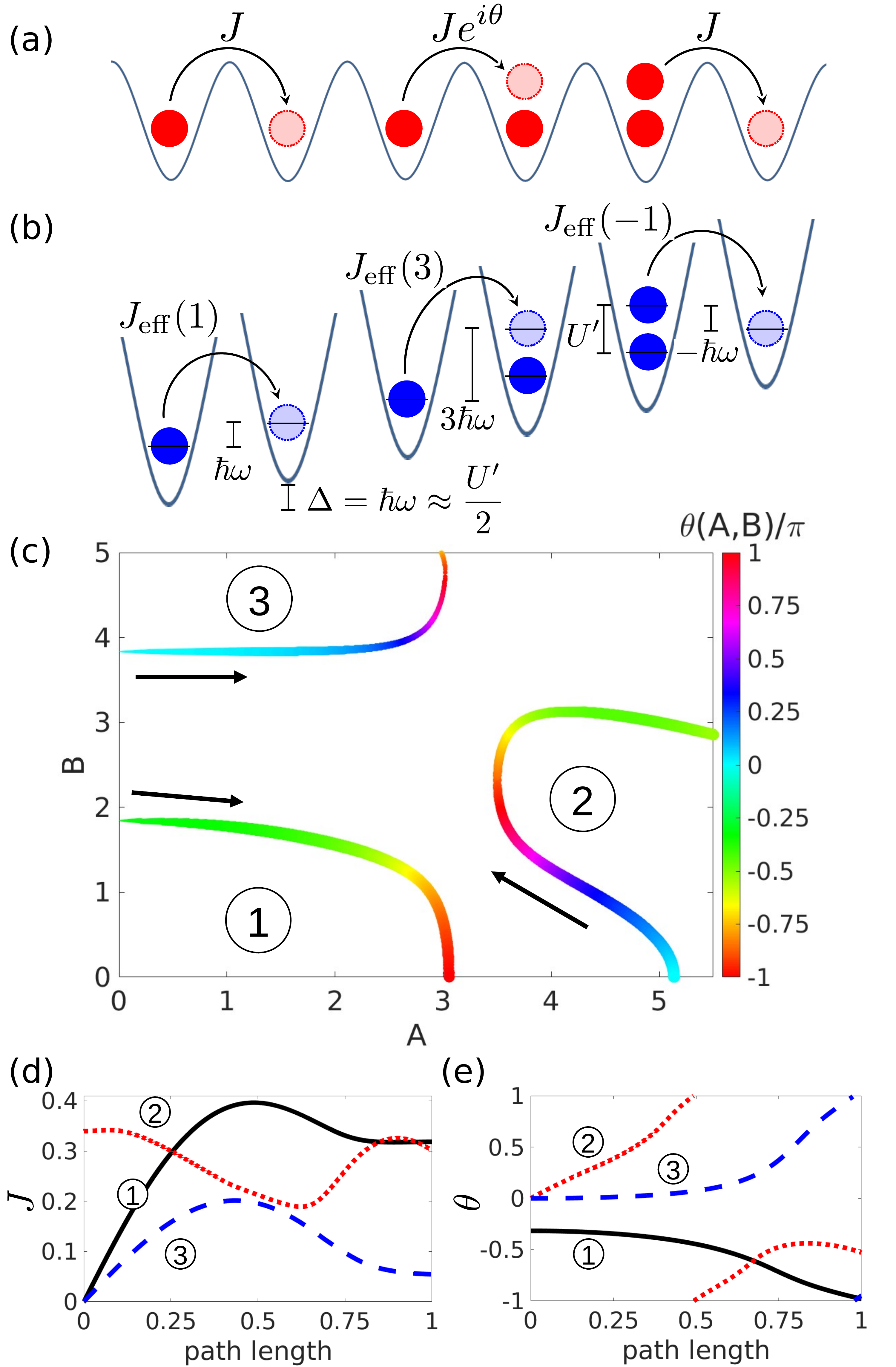}
\caption{(a) Basic number-dependent tunneling processes involving up to two bosons. We only depict rightwards tunneling, the leftwards processes are hermitian conjugated. (b) Realization as 1, 3 and -1 ``photon'' processes in a tilted lattice with strong on-site interactions $U'$ (b). We depict processes for tunneling rightwards; tunneling leftwards is simply described by conjugated processes. 
(c) Parameter curves that fulfill $|J_\text{eff}(1)|=|J_\text{eff}(3)|$. The color of the 
lines represents the statistical angle $\theta=\arg [J_\text{eff}(3)/J_\text{eff}(1)]$ and 
their thickness the tunneling amplitude $J=|J_\text{eff}(1)|$. (d,e) $\theta$ and $J$ 
following the lines in (b) in the direction of the arrow.}
\label{fig:figure1}
\end{figure}

The Hubbard model of one-dimensional lattice anyons with on-site interactions
\cite{KeilmannEtAl11} takes the form
\be
  \Ho = -J \sum_{j=2}^M \left(\hat{a}_j^{\dag} \hat{a}_{j-1} + \text{h.c.}\right) 
		+ U\sum_{j=1}^M \hat{n}_j\left(\hat{n}_j -1\right).
  \label{eq:ahm_any}
\ee
Here the annihilation and creation operators, $\ao_j$ and $\aa_j$, for anyons at site $j$ 
obey the commutation relations
$\hat{a}_j\hat{a}_k^{\dag} - e^{-i\theta \sgn(j-k)}\hat{a}_k^{\dag} \hat{a}_j =
\delta_{jk}$ and $\hat{a}_j\hat{a}_k - e^{-i\theta \sgn(j-k)}\hat{a}_k \hat{a}_j = 0$, 
which are parametrized by the statistical angle $\theta$. Here
$\sgn(k)=-1,0,1$ for $k<0,=0,>0$, respectively, so that on-site the particles behave 
like bosons. Thus even for $\theta=\pi$, these lattice anyons are just pseudofermions, since
many of them are allowed to occupy the same site. Following references
\cite{KunduEtAl99,KeilmannEtAl11}, the anyonic model can be mapped to the bosonic model
\bea
  \Ho = -J \sum_{j=2}^M \left(\hat{b}_j^{\dag} \hat{b}_{j-1} e^{i\theta\hat{n}_j}
		+ \text{h.c.}\right) + U\sum_{j=1}^M \hat{n}_j\left(\hat{n}_j -1\right).
  \label{eq:ahm_bos}
\eea
via the generalized Jordan-Wigner transformation
$\hat{a}_j = \hat{b}_j\ \exp\left( i\theta \sum_{k=j+1}^M \ba_k\bo_k\right)$.
Here the anyonic exchange phase has been translated to a density-dependent Peierls phase: 
when tunneling one site to the right (left), a boson pics up a phase given by $\theta$
($-\theta$) times the number of particles occupying the site it jumps to (from). Thus, if 
two particles pass each other via two subsequent tunneling processes to the right (left), 
the many-body wave function pics up a phase of $\theta$ ($-\theta$). These tunneling 
processes are illustrated in Fig.~\ref{fig:figure1}(a). 

For the realization of the number-dependent tunneling phase, we consider bosons 
in a tilted periodically forced lattice described by the Hamiltonian
\bea
\hat{H}(t) &=& \sum_j\Big(  -J^\prime \left[\hat{b}_j^\dagger \hat{b}_{j-1} + \mathrm{h.c.} \right] 
				+ \frac{U^\prime}{2} \hat{n}_j(\hat{n}_j-1)\nonumber\\
				 && + V_j\no_j
				 + \left[\Delta +  F(t)\right] j \hat{n}_j \Big).
\label{eq:H_time}
\eea
Here $J'>0$ and $U'>0$ denote the bare tunneling and interaction parameters, $\Delta>0$ 
characterizes a strong potential tilt, $V_j$ captures possible weak additional on-site
potentials, and $F(t)=F(t+T)$ incorporates a homogeneous time-periodic force of angular 
frequency $\omega=2\pi/T$ with vanishing cycle average $\frac{1}{T}\int_0^T\!\rd t\, F(t)=0$. It can be implemented as an inertial force
$F(t)/a =-m\ddot x(t)$, with lattice constant $a$, by shaking the lattice position
$x(t)$ back and forth. 
We require the resonance conditions
\bea
  \Delta = \hbar\omega,\quad U' = 2\hbar\omega + U,
\eea
as well as the high-frequency conditions
\be
J^\prime,|U|,|V_j-V_{j-1}| \ll \hbar\omega ,
\label{eq:high_freqi}
\ee
where we have introduced the small interaction detuning $U$. The largest share of the 
on-site energy is then given by $\Ho_0=\hbar\omega\sum_j[\no_j(\no_j-1)+j\no_j]$, so that
tunneling is energetically suppressed. Namely, when a particle tunnels from $j-1$ to $j$
this energy changes by 
$\hbar\omega\hat{\nu}_{j,j-1}$ with $\hat{\nu}_{j,j-1}= 2(\no_j -\no_{j-1}) + 3
=\pm\hbar\omega,\pm3\hbar\omega,\ldots$. 
However, coherent tunneling processes can be induced by time-periodic forcing as
$\nu$-``photon'' processes, where the drive provides or absorbs $|\nu|$ energy quanta
$\hbar\omega$. They are described by an effective tunneling matrix element
\cite{EckardtHolthaus07,SiasEtAl08}, which, through $\hat{\nu}_{j,j-1}$, will depend on 
the occupation numbers. Such number-dependent resonant tunneling has recently been 
investigated both experimentally \cite{MaEtAl11} and theoretically \cite{BermudezPorras15}.%
\footnote{An alternative approach for Floquet engineering number-dependent tunneling matrix 
elements relies on a modulation of the interaction strength \cite{GongEtAl09,RappEtAl12, 
GreschnerEtAl14, DiLibertoEtAl14, WangEtAl14, MeinertEtAl16}.} As we will show now, it can be 
used to achieve the number-dependent tunneling phases appearing in Eq.~(\ref{eq:ahm_bos}).

Using the time-periodic unitary operator 
\be\label{eq:U}
\Uo(t) = \exp\Big(-i \sum_j \Big[\omega t \, \no_j(\no_j -1 ) + \big\{\omega t -\chi(t)
\big\} j\no_j \Big]\Big),
\ee
where $\chi(t) = \frac{a m}{\hbar} \dot x(t)$ so that $\hbar\dot\chi(t)=-F(t)$, we can 
perform a number-dependent gauge transformation. It integrates out the strong on-site terms
$\Ho_0$ as well as the periodic force, but leads to number-dependent tunneling terms
$-J'\hat{b}_j^\dagger\hat{b}_{j-1}\exp[i\omega t \hat{\nu}_{j,j-1} - i\chi(t)]$ in the new 
Hamiltonian $\Uo^\dag(t)\Ho(t)\Uo(t)-i\hbar\Uo^\dag(t)\partial_t\Uo(t)$. Averaging over
the rapidly oscillating phase factor by integrating over one driving period  
(corresponding to the leading order of a high-frequency approximation
\cite{GoldmanDalibard14, BukovEtAl15, EckardtAnisimovas15}), we obtain the effective
time-independent Hamiltonian 
\bea
  \Ho_{\mathrm{eff}}
	&=& - \sum_j \left( \hat{b}_j^\dagger \hat{b}_{j-1} 
			J_{\mathrm{eff}}(\hat{\nu}_{j,j-1})  + \mathrm{h.c.} \right)\nonumber\\
		 && +\sum_j \left(\frac{U}{2} \hat{n}_j(\hat{n}_j-1)  + V_j  \hat{n}_j \right).
\label{eq:H_eff}
\eea
It contains the number-dependent tunneling parameter
\be\label{eq:Jeff}
J_{\mathrm{eff}}(\nu) 
=  \frac{J'}{T} \int_0^T\!\rd t \, \exp\Big(i\omega t \nu - i\chi(t)\Big) 
\ee
and the tunable interaction parameter $U=U'-2\hbar\omega$, which can take both
negative and positive values. 

The effective tunneling matrix elements $J_{\mathrm{eff}}(\nu)$ should reproduce the
number-dependent tunneling parameters of Eq.~(\ref{eq:ahm_bos}). We restrict ourselves to 
the low-density regime, where the dominant processes involve one or two bosons. These 
processes are those depicted in Fig.~\ref{fig:figure1}(a) as well as the hermitian
conjugated processes for tunneling leftwards. As illustrated in Fig.~\ref{fig:figure1}(b),
these processes are associated with different potential energy changes
$\nu\hbar\omega$. Tunneling rightwards from a singly or doubly occupied site onto an empty 
site corresponds to $\nu=1$ or $\nu=-1$, respectively, and should be described by the 
parameter,
\be
J_{\mathrm{eff}}(1) = J_{\mathrm{eff}}(-1) = Je^{i\phi_g} ,  
\label{eq:J2}
\ee
with real tunneling amplitude $J$ and arbitrary Peierls phase $\phi_g$ reflecting the 
freedom of gauge. Tunneling rightwards from an empty site onto an occupied site is 
associated with $\nu=3$ and the corresponding tunneling parameter should carry an 
additional phase $\theta$, 
\be
J_{\mathrm{eff}}(3) =  Je^{i\theta+i\phi_g}.
\label{eq:J1}
\ee

In order to fulfill conditions (\ref{eq:J2}) and (\ref{eq:J1}), we make the simple ansatz
\be\label{eq:ansatz}
\chi(t) = A\cos(\omega t) + B\cos(2\omega t)
\ee
for the (integrated) driving force (other choices are possible). This ansatz already 
ensures that $J_{\mathrm{eff}}(1)=J_{\mathrm{eff}}(-1)$. The additional constraint
$|\J_{\mathrm{eff}}(3)|=J=|J_\text{eff}(1)|$ defines lines in the $A$-$B$ plane, as 
can be seen in panel (c) of Fig.~\ref{fig:figure1}.  The thickness and the color of the plotted 
lines represents the tunneling amplitude $J$ and the statistical angle $\theta$, respectively. 
The variation of $J$ and $\theta$ along the lines is also plotted in panels (d) and (e). Whereas 
lines 2 and 3 cover the full range $|\theta|\in[0,\pi]$, line 1 roughly allows to realize
$|\theta|\in [0.4\pi,\pi]$.

\begin{figure}[t]
\centering
\includegraphics[width=0.5\textwidth,]{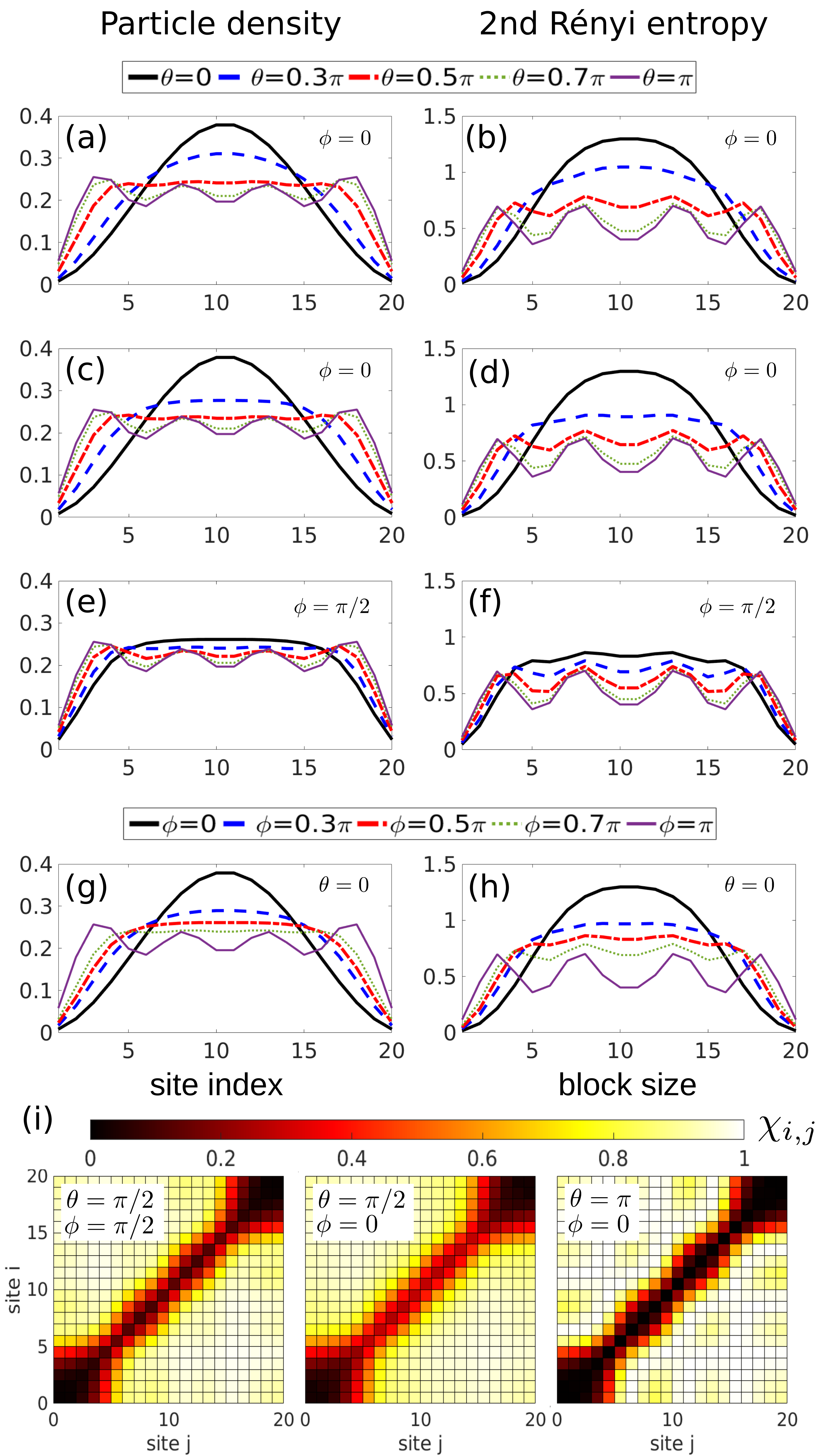} \hspace{-0.5cm}
\caption{Ground-state properties: Density $n_i$ and entanglement entropy $S_\ell$
of the first $\ell$ sites. Either the anyonic angle $\theta$ is 
varied (a-f) or the relative interaction strength $U/J=\tan(\phi/2)$ (g,h). (i) 
Computed for the effective Hamiltonian (\ref{eq:H_eff}) (a,b), 
the ideal model (\ref{eq:ahm_bos}) (c-f), plain bosons (g,h). Two-particle correlation function
$\chi_{i,j} = \langle b_i^\dag b_j^\dag b_j b_i\rangle/(n_i n_j)$ for various anyonic
angles $\theta$ and interaction strengths $U/J=\tan(\phi/2)$ for the effective Hamiltonian 
(\ref{eq:H_eff}).}
\label{fig:figure2}
\end{figure}

A clear signature of the continuous fermionization of the system with increasing $\theta$ 
is the formation of a Fermi sea in the momentum distribution of the anyons (see e.g.\ 
Ref.~\cite{TangEtAl15}). However, in an experiment one cannot measure the anyonic momentum 
distribution, but only the bosonic one, which, due to both the Jordan-Wigner transformation 
and the gauge transformation (\ref{eq:U}) differs from that of the anyons. Therefore, in the 
following we will consider only such observables that are invariant under these 
transformations. These include the densities $n_i=\la\no_i\ra$, the two-particle correlations
$\chi_{i,j} = \langle b_i^\dag b_j^\dag b_j b_i\rangle/(n_i n_j)$, and also the second
R\'enyi entropy characterizing the purity of the reduced density matrix $\hat{\rho}_\ell$ of 
the subsystem given by the first $\ell$ sites $j=1,\ldots, \ell$, 
$ S_\ell = -\ln\mathrm{Tr}(\hat{\rho}_\ell^2)$.\footnote{
The Fock states $|{\bm n}\ra_c = |n_1\ra_c|n_2\ra_c\cdots|n_M\ra_c $ for 
1D anyons ($c=a$) and bosons ($c=b$) obey the Jordan-Wigner transformation
$|{\bm n}\ra_a=\exp(-i\theta\sum_{j=1}^M\sum_{k=j+1}^M n_k)|{\bm n}\ra_b$. It corresponds to 
independent site-local transformations $|n_j\ra_a=\exp(-i\theta(j-1)n_j)|n_j\ra_b$, so that 
tracing out a site $k$ is identical for bosons and 1D anyons, 
$\sum_{n_k} {}_a\la n_k|\cdot|n_k\ra_a =\sum_{n_k} {}_b\la n_k|\cdot|n_k\ra_b$.} 
In the following, we will focus on ground-state properties, so that $S_\ell$ is an entanglement 
entropy (as it has been measured recently in a bosonic chain \cite{IslamEtAl15}).

We compute $n_i$, $\chi_{i,j}$, and $S_\ell$ using exact diagonalization both for the 
ideal model (\ref{eq:ahm_bos}) and the effective Hamiltonian (\ref{eq:H_eff}). We 
consider $N=4$ bosons on $M=20$ sites, corresponding to a density of $n=0.2$. The 
effective tunneling matrix elements (\ref{eq:Jeff}) were obtained for the driving 
function corresponding either to path 1 of Fig.~\ref{fig:figure1}(c) or to path 2, in 
case the desired statistical angle $|\theta|$ is not available in path 1. Despite the 
fact that they reproduce the ideal tunneling matrix elements only for processes involving 
one and two particles, we find very good agreement between the ideal and the effective 
model: In Figs.~\ref{fig:figure2}(a) and (b), we plot $n_i$ and $S_\ell$ for the effective 
model with $U=0$ and various anyonic angles $\theta$. The results match very well with 
those obtained for the ideal model shown in Figs.~\ref{fig:figure2}(c) and (d). 
For non-zero on-site interactions, $U/J\equiv\tan(\phi/2)$ the agreement is equally good, so 
that we only plot the results for the ideal model in Fig.~\ref{fig:figure2}(e) and (f).

In Fig.~\ref{fig:figure2}(c), we can observe that the density distribution flattens in 
the center, when the statistical angle is switched on. This effect can be understood, by 
noting that the scattering properties resulting from the density dependent tunneling 
resemble those of repulsive on-site interactions \cite{GreschnerSantos15}, which favor a 
flat density. For large $\theta$, the density becomes modulated, with one maximum for 
each particle in the system. These oscillations correspond to Friedel oscillations, which 
are a hallmark of fermionic behavior \cite{Friedel58}. They are a finite-size effect 
induced by the hard-wall boundary conditions (as they can be realized in quantum-gas 
microscopes). Their wavelength is roughly given by $\pi/k_F$, with Fermi wave number $k_F$,
corresponding to the average particle distance, which in our system is given by $n^{-1}=5$
lattice constants. Generally, Friedel oscillations occur in the vicinity of localized 
defects. Their build-up allows us to monitor the continuous fermionization of the 1D 
anyons in a system of bosons with number-dependent tunneling. The oscillations are also 
clearly visible in the entanglement entropy [Fig.~\ref{fig:figure2}(d)]. An intuitive 
explanation is that the maximum corresponds to the position of a particle, whose 
delocalization within the maximum contributes to the entanglement between the left and 
right subsystems.

\begin{figure}[t]
\centering
\includegraphics[width=0.5\textwidth,]{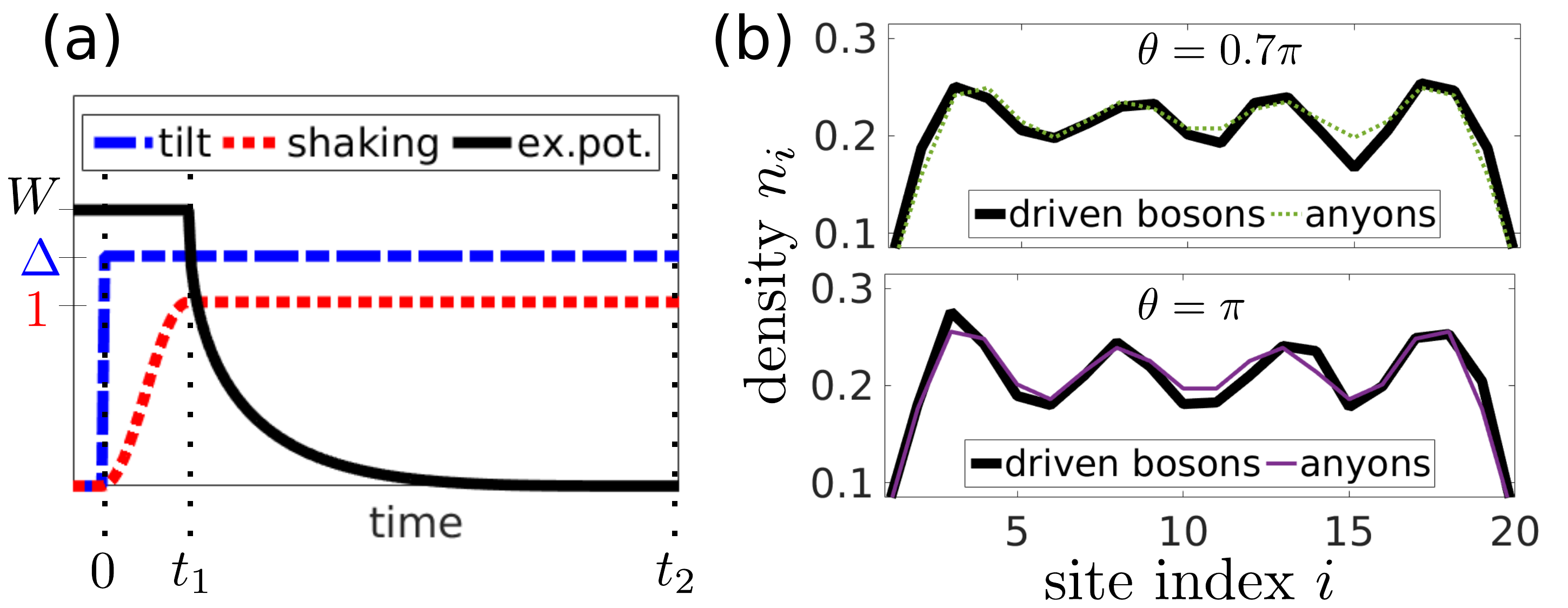} \hspace{-0.5cm}
\caption{State preparation for a system of 4 particles on 20 sites, simulated 
using the full time-dependent Hamiltonian (\ref{eq:H_time}). (a) 
Preparational protocol. (b) Density distribution of the final state compared 
to the ground state of the anyon model (\ref{eq:ahm_any}). We have chosen 
realistic parameters for an optical lattice of depth $V_0=10E_R$ giving
$J^\prime =0.0192E_R$, where the recoil energy $E_R$ typically corresponds to 
frequencies of a few kilo Hertz \cite{BlochDalibardZwerger08}. Moreover, we 
chose $\hbar\omega=E_R-U$ (well below the band gap of $\approx5 E_R$),
$U=0.5J$, $W= 0.6 E_R$, $t_1=50 T$ and $t_2= 1240T (1300T)\approx 50\hbar/J$ 
for $\theta=0.7\pi (\pi)$.} 
\label{fig:figure3}
\end{figure}

In Figs.~\ref{fig:figure2}(e,f), showing data for a system with significant
on-site interactions $U/J=\tan(\pi/4)=1$, the fermionic signatures occur 
already for smaller $\theta$. This observation is consistent with the
well-known fact that increasing the on-site interactions is another way of 
approaching fermionic behavior \cite{ParedesEtAl04}. This is illustrated also 
in Fig.~\ref{fig:figure2}(g,h) showing data for plain bosons ($\theta=0$) and 
different $U/J$. In the hard-core limit ($\phi=\pi$), the bosons can be 
mapped to fermions. Despite the fact that 1D anyons only become 
pseudofermions for $\theta=\pi$, the data for $\theta=\pi$ in
Figs.~\ref{fig:figure2}(c,d) agrees very well with that for $\phi=\pi$ in 
Figs.~\ref{fig:figure2}(g,h). This suggests that for low densities 
pseudofermions behave like true fermions. This is confirmed also by the 
correlations shown in Fig.~\ref{fig:figure2}(i). Their diagonal elements 
$\chi_{j,j}=\la\no_j(\no_j-1)\ra/n_j^2$, which are a measure for double 
occupation, vanish for $\theta=\pi$, even though pseudofermions locally 
behave like bosons. Simulations show that pseudofermions behave like 
fermions up to a filling of about one third \cite{supplemental}. This is also 
the filling, where the low-density description, stating that driven bosons 
behave like anyons, is found to break down \cite{supplemental}.

The build up of Friedel oscillations requires the system to be quantum 
degenerate, i.e.\ temperatures $\mathcal{T}$ well below the Fermi energy
$E_F=2J[1-\cos(k_F)]\approx J\pi^2 n^2$, so that the thermal wavelength
is large compared to the mean particle distance. Computing canonical
expectation values for $N=4$ anyons on $M=20$ sites for the finite
temperature $\mathcal{T}/J=0.1$ (corresponding to an von Neumann entropy
per particle of $s\approx 0.20$, which is accessible in a system of
spinless bosons), we find well pronounced Friedel oscillations \cite{supplemental}. In an
experiment, the low-entropy state of the effective Hamiltonian
(\ref{eq:H_eff}) has to be prepared starting from a low-entropy state of
the undriven system. Let us assume that initially $\Delta=F=0$ and the
system is prepared in a Mott-insulator state 
$|S\ra=\prod_{j\in S}\ba_j|\text{vac}\ra$, with the set $S$ containing $N$
lattice sites and $|\text{vac}\ra$ denoting the vacuum state. This is the
asymptotic ground state in the presence of an external potential
$V_j=-W\delta_{j\in S}$ in the limit $W,U'\gg J$. For finite $U$, this is
also the ground state of the effective model (\ref{eq:H_eff}), with
$J_\text{eff}(\nu)=0$. Thus, we can adiabatically melt the Mott insulator
into the ground state of $\Ho_\text{eff}$ by smoothly ramping up the forcing,
i.e.\ the $J_\text{eff}(\nu)$, and then continuously switching off the
external potential $W$. In order to minimize the mass transport during this
adiabatic process, it is useful that $S$ contains equally spaced lattice sites,
so that $V_i$ describes a superlattice. We have simulated this protocol
integrating the time evolution of the full time-dependent Hamiltonian
(\ref{eq:H_time}) and find excellent agreement between the final state and
the ground state of $H_\text{eff}$ [Fig.~\ref{fig:figure3}]. This confirms
both a description in term of the effective Hamiltonian and the proposed 
preparation scheme.

In summary, we proposed a simple scheme for the realization of 1D anyons, 
which is feasible in existing experimental setups. It is based on Floquet 
engineering a system of bosonic atoms with number-dependent tunneling phases. 
We, moreover, showed that Friedel oscillations can serve as a directly 
measurable signature for the continuous fermionization of the anyons.

\begin{acknowledgments}

We thank Sebastian Greschner, Axel Pelster, and Luis Santos for useful discussions. CS is 
grateful for support by the Studienstiftung des deutschen Volkes. This work was supported by
the DFG via the Research Unit (Forschergruppe) FOR 2414. 

\end{acknowledgments}

\section*{Supplemental Material}
\subsection*{Behavior with Respect to Density}

\begin{figure}[h]
\includegraphics[width=1\linewidth]{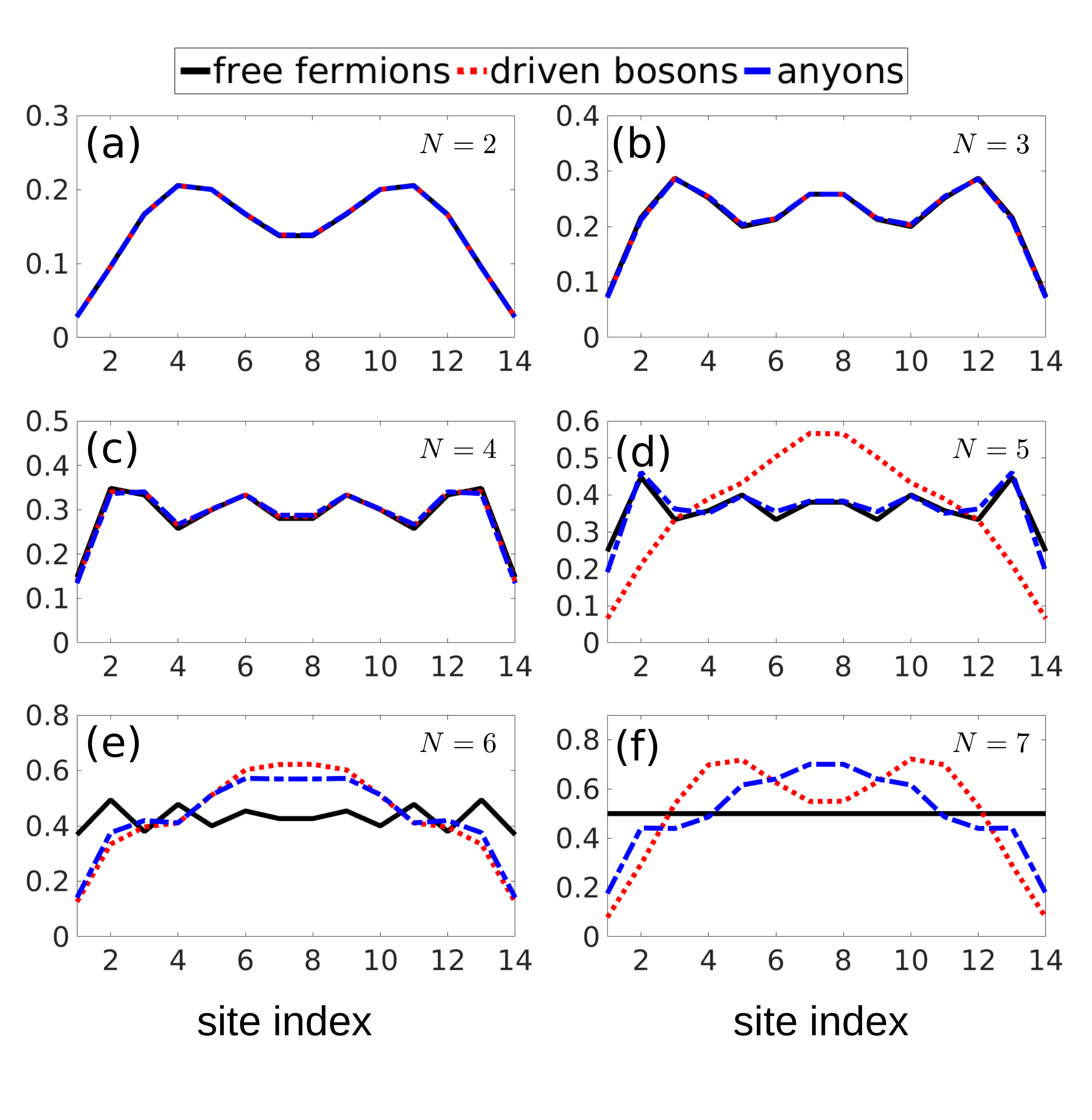}
\centering
\caption{On-site densities $n_i$ computed for the ground state of $N$ 
particles on an open chain of $M=14$ sites, for 1D anyons with $\theta=\pi$
(dashed blue lines), bosons described by the effective Hamiltonian (7)
(dotted red lines), and free fermions (solid black line).} 
\label{fig:figure4}
\end{figure}

Let us first explore how the system behaves when the filling is increased. 
In Fig.~\ref{fig:figure4}, we plot the ground-state density profile for a 
small chain of $M=14$ sites with open boundary conditions for various 
particle numbers between $N=2$ and $N=7$. We compare results for 1D anyons
(dashed blue lines) with $\theta = \pi$, which are pseudofermions that behave 
on site like bosons, with results obtained for both free fermions (solid 
black lines) and bosons described by the effective Hamiltonian (7) given in 
the main text (dotted red lines). 

The effective bosonic Hamiltonian is constructed to reproduce the anyonic 
behavior for single- and two-particle processes. Accordingly, its ground state 
is expected to mimic that of the anyonic model for low densities. The 
simulations presented in Fig.~\ref{fig:figure4}, which are based on exact 
diagonalization, confirm this behavior. We find almost perfect agreement 
between bosons and anyons, until the low-density approximation breaks down 
abruptly at a filling of $n=5/14\approx0.36$. 

In Figure 2(i) of the main text, we can observe that the probability for 
finding two pseudofermions on the same site practically vanishes like for 
actual fermions. The results presented in Fig.~\ref{fig:figure4} show that 
pseudofermions mimic almost perfectly the behavior of free fermions for low 
densities, up to a filling of $n=5/14$. Here the fermionic density is given by
\be
n_j^\text{fermions} = \sum_{\alpha=1}^N |\psi^{(\alpha)}_j|^2,
\ee
where  
\bea
  \psi^{(\alpha)}_j =  \sqrt{\frac{2}{M+1}} \sin(k_\alpha j),
\eea
are the wave functions of the  single-particle eigenstates
$\alpha=1,2,\ldots,M$ with wave numbers $k_\alpha=\alpha 2\pi/(M+1)$ and 
energies $E_\alpha=-2J[1-\cos(k_\alpha)]$. 

An intuitive explanation for the excellent agreement between pseudofermions 
and fermions at low densities can be given as follows. The number-dependent 
tunneling terms appearing in the bosonic representation (2) of the anyonic 
model (1) give rise to two-particle scattering, which is described by a 
scattering length that diverges for $\theta=\pi$ (see Ref.~[35] of the main 
text). Thus, in the low-density regime, which is governed by two-particle 
processes, we recover the physics of hard-core bosons, which in turn can be 
mapped to (actual) fermions.

\subsection*{Finite-Temperature Behavior}
 
\begin{figure}[h]
\includegraphics[width=1\linewidth]{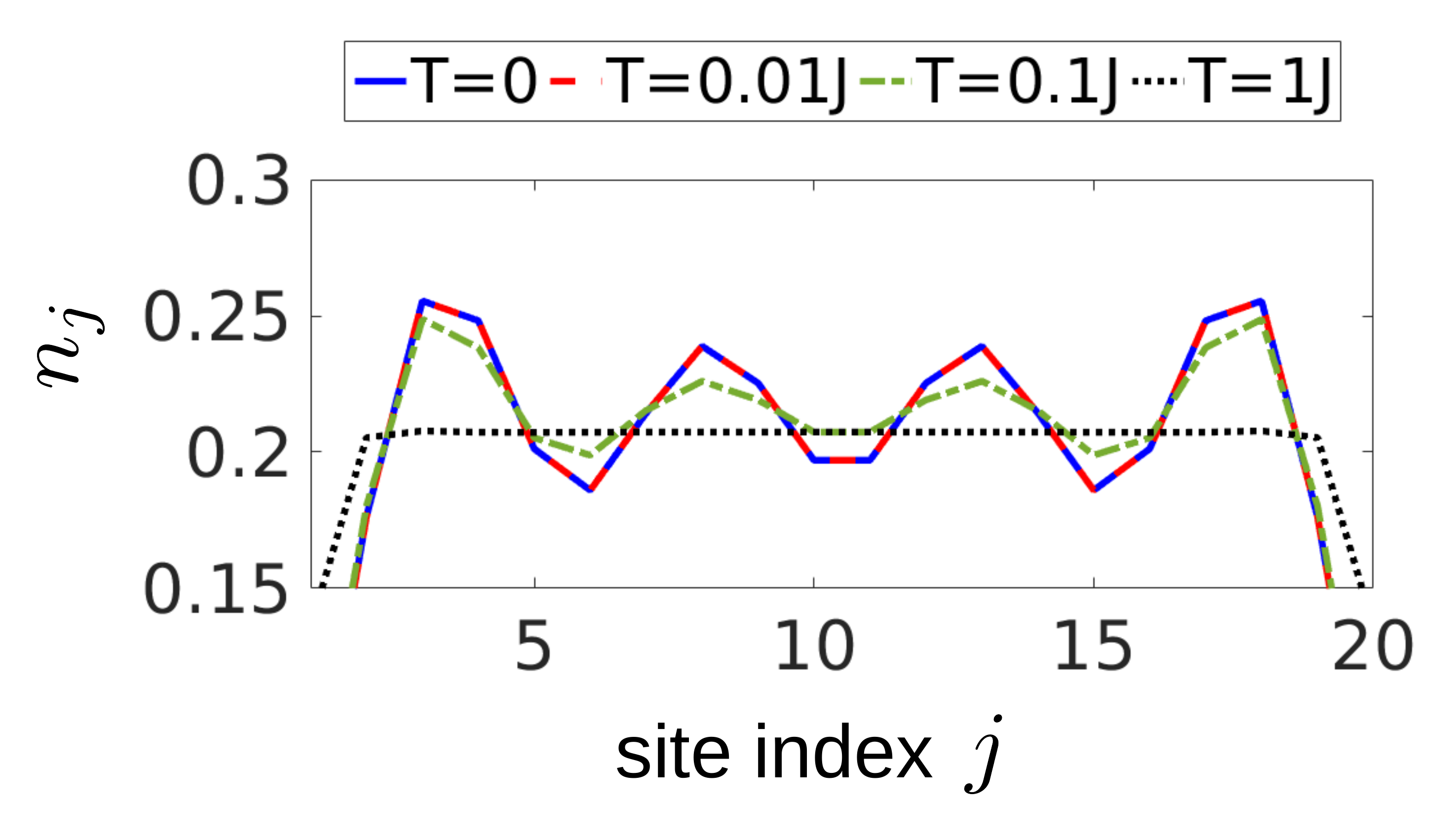}
\centering
\caption{Density profile for the anyonic model with $N=4$ particles on $M=20$ sites,
$U/J=0.5$, and temperatures $\mathcal{T}/J=0$, $0.01$, $0.1$, $1$.}
\label{fig:figure5}
\end{figure}

Let us finally investigate the effect of a finite temperature. For a system of $N=4$ 
anyons on $M=20$ lattice sites with $U/J=0.5$, we compute the canonical expectation value 
for the density profile by using exact diagonalization for computing not only the ground 
state, but also the excited states of the interacting system. We compare results obtained 
for four different temperatures $\mathcal{T}$. The corresponding (von Neumann) entropies per 
particle are calculated to be given by $s=0$, $1.1\cdot 10^{-7}$, $0.20$, and $1.6$ (in 
units of the Boltzmann constant), respectively. Figure \ref{fig:figure5} shows how the 
density profile of the anyonic model changes when the temperature is increased. One can 
clearly see that the Friedel oscillations are preserved as long as $\mathcal{T}\ll J$.

\bibliography{mybib}

\end{document}